\documentclass[twocolumn,showpacs,preprintnumbers,amsmath,amssymb,prl]{revtex4-1}
\usepackage{mathrsfs}
\usepackage{graphicx}
\usepackage{dcolumn}
\usepackage{bm}
\usepackage{amsmath}
\usepackage{amsfonts}

\begin{document}

\title{Quantum Anomalous Hall Effect with Higher Plateaus}
\author{Jing Wang}
\author{Biao Lian}
\author{Haijun Zhang}
\author{Yong Xu}
\author{Shou-Cheng Zhang}
\affiliation{Department of Physics, McCullough Building, Stanford University, Stanford, California 94305-4045, USA}

\begin{abstract}
Quantum anomalous Hall (QAH) effect in magnetic topological insulators
is driven by the combination of spontaneous magnetic moments and spin-orbit
coupling. Its recent experimental discovery raises the question if higher plateaus can also be realized.
Here we present a general theory for QAH effect with higher Chern numbers, and
show by first-principles calculations that thin film magnetic topological insulator of
Cr-doped Bi$_2$(Se,Te)$_3$ is a candidate for the $C=2$ QAH insulator. Remarkably,
whereas higher magnetic field leads to lower Hall conductance plateaus in the integer
quantum Hall effect, higher magnetic moment leads to higher Hall conductance plateaus
in the QAH effect.
\end{abstract}

\date{\today}


\maketitle

The topological phases of two-dimensional (2D) insulators with broken time reversal symmetry
is characterized by the first Chern number~\cite{thouless1982}, which takes integer values in the
integer quantum Hall effect (IQHE). In the IQHE, electronic states of 2D electron system form Landau
levels under strong external magnetic fields, and the Hall resistance is quantized into $h/Ce^2$
plateaus~\cite{klitzing1980} contributed by dissipationless chiral states at sample
edges~\cite{halperin1982} (where $h$ is Plank's constant, $e$ is the charge of an electron, and $C$ is the Chern number).
In principle, quantum Hall effect can exist without the external magnetic field and the associated
Landau levels~\cite{haldane1988}, however, the Haldane model~\cite{haldane1988} with circulating currents on a honeycomb
lattice is not easy to implement experimentally.
In a QAH insulator, theoretically proposed for magnetic topological insulators (TIs)~\cite{qi2006,qi2008,liu2008,li2010,yu2010},
the ferromagnetic (FM) ordering and spin-orbit coupling (SOC) are sufficiently strong that they can give rise to a topologically nontrivial
phase with finite Chern number. Recently, the QAH effect has been experimentally discovered in magnetic
TI of Cr-doped (Bi,Sb)$_2$Te$_3$, where the $C=1$ has been reached~\cite{chang2013b}.
Search for QAH insulator with higher Chern numbers could be important both for fundamental and practical
interests. The edge channels of the QAH insulator has been proposed as interconnects for integrated circuits~\cite{zhang2012}.
However, while the edge channels of the QAH insulator conducts without dissipation, contact resistance
could still limit possible application in interconnects. QAH effect with higher plateaus lowers the
contact resistance, significantly improving the performance of the interconnect devices.
Fractional filling of Chern insulators with $C=2$ could also lead to new topological states with novel
elementary excitations~\cite{maissam2012}. QAH effect with higher plateaus also shows dramatic difference between
the IQHE and the QAH effect: whereas higher magnetic field leads to lower Hall conductance plateaus in the IQHE, higher magnetic moment leads to higher Hall conductance plateaus in the QAH effect.

In this Letter, we present a general theory for QAH effect with higher plateaus. Based on the first-principles calculations,
we predict that thin films of Cr-doped Bi$_2$(Se$_x$Te$_{1-x}$)$_3$ TI is a candidate for the $C=2$ QAH insulator. The
tunable magnetic ordering and SOC in this system provide an ideal platform for realizing other exotic topological states in magnetic TIs.

The basic mechanism for the quantum spin Hall effect or TI is the band inversion of
spin degenerate bands, described by the Bernevig-Hughes-Zhang model~\cite{bernevig2006c}.
Similarly, the basic mechanism for the QAH effect is the band inversion of spin polarized
bands in magnetic TIs~\cite{qi2006,qi2008,liu2008,li2010,yu2010}. The general theory for higher Chern number
QAH effect presented in this letter is generic for any thin films of magnetic TIs. We would like to start
from a simple model describing TIs Bi$_2$Te$_3$, Bi$_2$Se$_3$ and Sb$_2$Te$_3$ for concreteness~\cite{zhang2009}.
The thin films made out of this family of compounds doped with Cr or Fe develops ferromagnetism even up to $190$~K~\cite{kulbachinskii2001,zhou2006,chang2013a}. The QAH effect can be realized in 2D thin film of such magnetic TIs with spontaneous FM order. The low-energy bands of these materials consist of a
bonding and an antibonding state of $p_z$ orbitals, labelled by
$\left|P2^-_z,\uparrow(\downarrow)\right\rangle$ and $\left|P1^+_z,\uparrow(\downarrow)\right\rangle$, respectively. The generic form of the effective Hamiltonian describing these four bands is
\begin{eqnarray}\label{model}
\mathcal{H}_{\mathrm{3D}}(k_x,k_y,k_z)
   &=&\left(\begin{array}{cc}
   H_+(k) & A_1k_zi\sigma_y\\
   -A_1k_zi\sigma_y & H_-(k)
   \end{array}\right),\\
H_{\pm}(k) &=& \varepsilon(k)+d^i_{\pm}(k)\tau_i,
\end{eqnarray}
here $\tau_i$ ($i=1,2,3$) and $\sigma_y$ are Pauli matrices. $d^{1,2,3}_{\pm}(k)=(A_2k_x,\pm A_2k_y,M(k)\mp\Delta)$. To the lowest order in $k$, $M(k)=B_0+B_1k_z^2+B_2(k_x^2+k_y^2)$, $\varepsilon(k)=D_0+D_1k_z^2+D_2(k_x^2+k_y^2)$ accounts for the particle-hole asymmetry. $B_0<0$ and $B_1, B_2>0$ guarantee the system is in the inverted regime. The basis of Eq.~(\ref{model}) is $|P1^{+}_z,\uparrow\rangle$, $|P2^{-}_z,\downarrow\rangle$, $|P1^{+}_z,\downarrow\rangle$, $|P2^{-}_z,\uparrow\rangle$, and the $\pm$ in the basis stand for the even and odd parity and $\uparrow$, $\downarrow$ represent spin up and down states, respectively. $\Delta$ is the exchange field along the $z$ axis introduced by the FM ordering. For simplicity, the same effective $g$-factor for the two oribtals $P1^+_z$ and $P2^-_z$ is assumed.

\begin{figure}[t]
\begin{center}
\includegraphics[width=3.4in,clip=true]{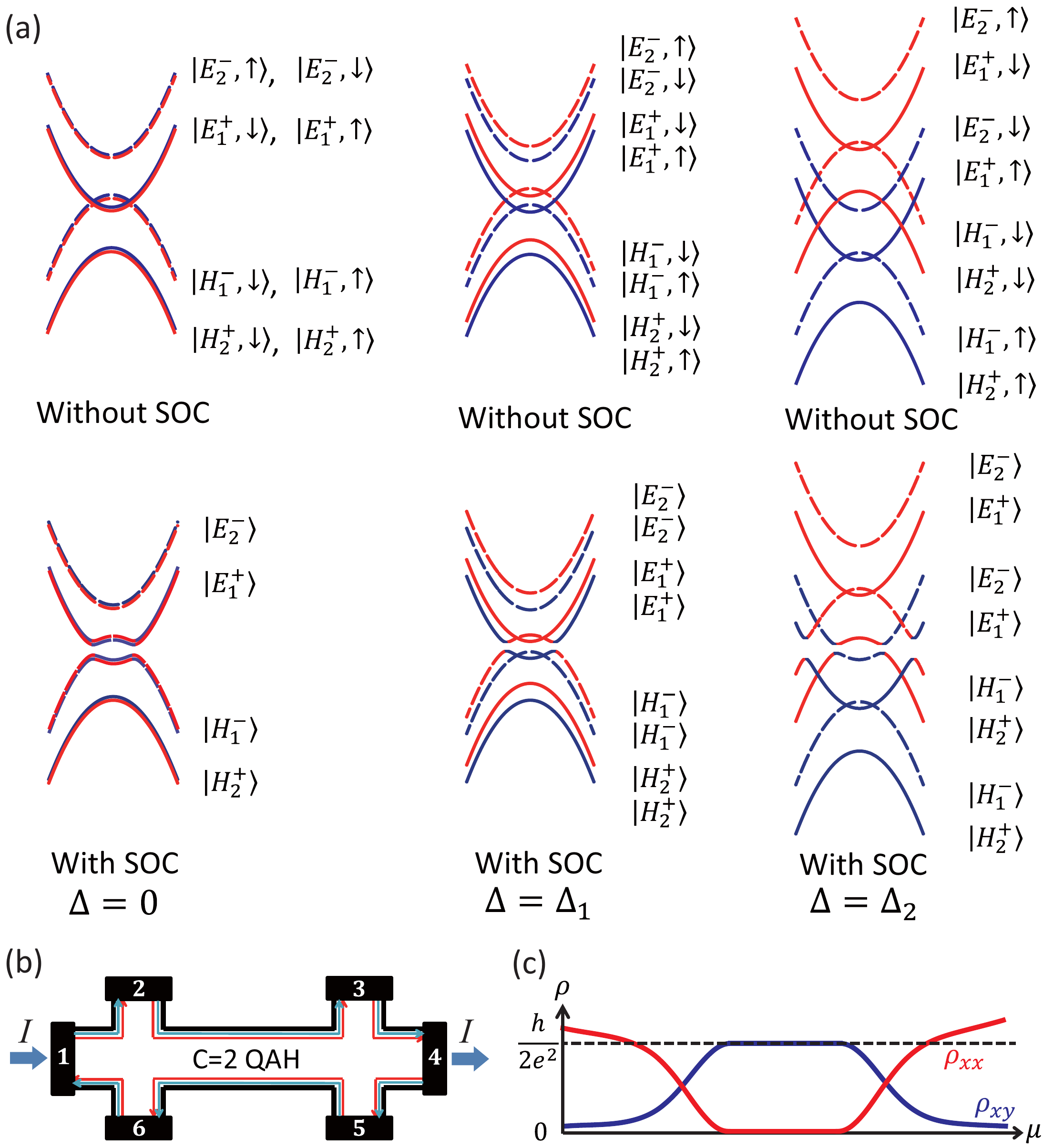}
\end{center}
\caption{Evolution of the subband structure upon increasing the exchange field. The solid lines denote the sub-bands that have even parity at $\Gamma$ point, and dashed lines denote sub-bands with odd parity at $\Gamma$ point. The blue color denotes the spin up electrons; red, spin down electrons. (a) The initial (E$_1$, H$_1$) sub-bands are already inverted, while the (E$_2$, H$_2$) subbands are not inverted. The exchange field $\Delta_1$ release the band inversion in one pair of (E$_1$, H$_1$) subbands and increase the band inversion in the other pair, while the (E$_2$, H$_2$) subbands are still not inverted. With stronger exchange field $\Delta_2$, a pair of inverted (E$_2$, H$_2$) subbands appears, while keeping only one pair of (E$_1$, H$_1$) subbands inverted. (b) Schematic drawing of a Hall bar device of $C=2$ QAH effect, and (c) expected chemical potential dependence of zero magnetic field $\rho_{xx}$ (in red) and $\rho_{xy}$ (in blue).} \label{fig1}
\end{figure}

The confinement of thin films of three-dimensional (3D) magnetic TIs in the $z$ direction quantizes the momentum on this axis and
leads to 2D sub-bands labeled by the sub-band index $n$.
In order to illustrate the underlying physics clearly, we first take the limit $A_1=0$, in which case the
system is decoupled into two classes of 2D models $h_+(n)$ and $h_-(n)$ with opposite chirality
\begin{equation}
\tilde{\mathcal{H}}_{\mathrm{2D}}(n)
  =\left(\begin{array}{cc}
  h_+(n) & 0\\
  0 & h_-(n)
  \end{array}\right)
\end{equation}
where $h_{\pm}(n)=\tilde{\varepsilon}_n1_{2\times2}+(\tilde{M}_n\mp\Delta)\tau_3+A_2k_x\tau_1\pm A_2k_y\tau_2$, expressed in the subspace of $|E_n,\uparrow\rangle=\varphi_n(z)|P1_z^+,\uparrow\rangle$, $|H_n,\downarrow\rangle=\varphi_n(z)|P2_z^-,\downarrow\rangle$ for $h_+(n)$ and $|E_n,\downarrow\rangle=\varphi_n(z)|P1_z^+,\downarrow\rangle$, $|H_n,\uparrow\rangle=\varphi_n(z)|P2_z^-,\uparrow\rangle$ for $h_-(n)$. $\tilde{\varepsilon}_n=D_0+D_1\langle k_z^2\rangle_n+D_2(k_x^2+k_y^2)$, $\tilde{M}_n=B_0+B_1\langle k_z^2\rangle_n+B_2(k_x^2+k_y^2)$,  and the confinement in a thin film of thickness $d$ is given by the relation $\varphi_n(z)=\sqrt{2/d}\sin(n\pi z/d+n\pi/2)$ and $\langle k_z^2\rangle_n=(n\pi/d)^2$ for sub-bands index $n$. $|E_n,\uparrow\backslash\downarrow\rangle$ and $|H_n,\uparrow\backslash\downarrow\rangle$ have parity $(-1)^{n+1}$ and $(-1)^n$, respectively. At half filling, the effective models $h_{\pm}(n)$ have Chern number $\pm 1$ or $\pm 0$ depending on whether the Dirac mass is inverted ($\tilde{M}_n\mp\Delta<0$) or not ($\tilde{M}_n\mp\Delta>0$) at $\Gamma$ point. Thus the total Chern number of the system is
\begin{equation}\label{chern}
C = N_+-N_-
\end{equation}
where $N_{\pm}$ is the number of $h_{\pm}(n)$ with inverted Dirac mass, respectively. As shown in Fig.~\ref{fig1}(a), when $\Delta=0$, $N_+=N_-$, thus the net Hall conductance of this system vanishes; while the $Z_2$ index $N_+(\mathrm{mod~2})$, can be still nonzero, which gives the crossover from 3D TI to quantum spin Hall insulator~\cite{liu2010a}. When $\Delta\neq0$, $N_+$ can be different from $N_-$. In the $\Delta=\Delta_1$ case, only the Dirac mass of $h_+(1)$ is inverted, thus $N_+=1$ and $N_-=0$, the system is in a QAH state with $C=1$. When the exchange field is larger with $\Delta=\Delta_2$, the Dirac mass of $h_+(1)$ and $h_+(2)$ are inverted, $N_+=2$ and $N_-=0$, gives QAH phase with $C=2$.

\begin{figure}[t]
\begin{center}
\includegraphics[width=3.3in]{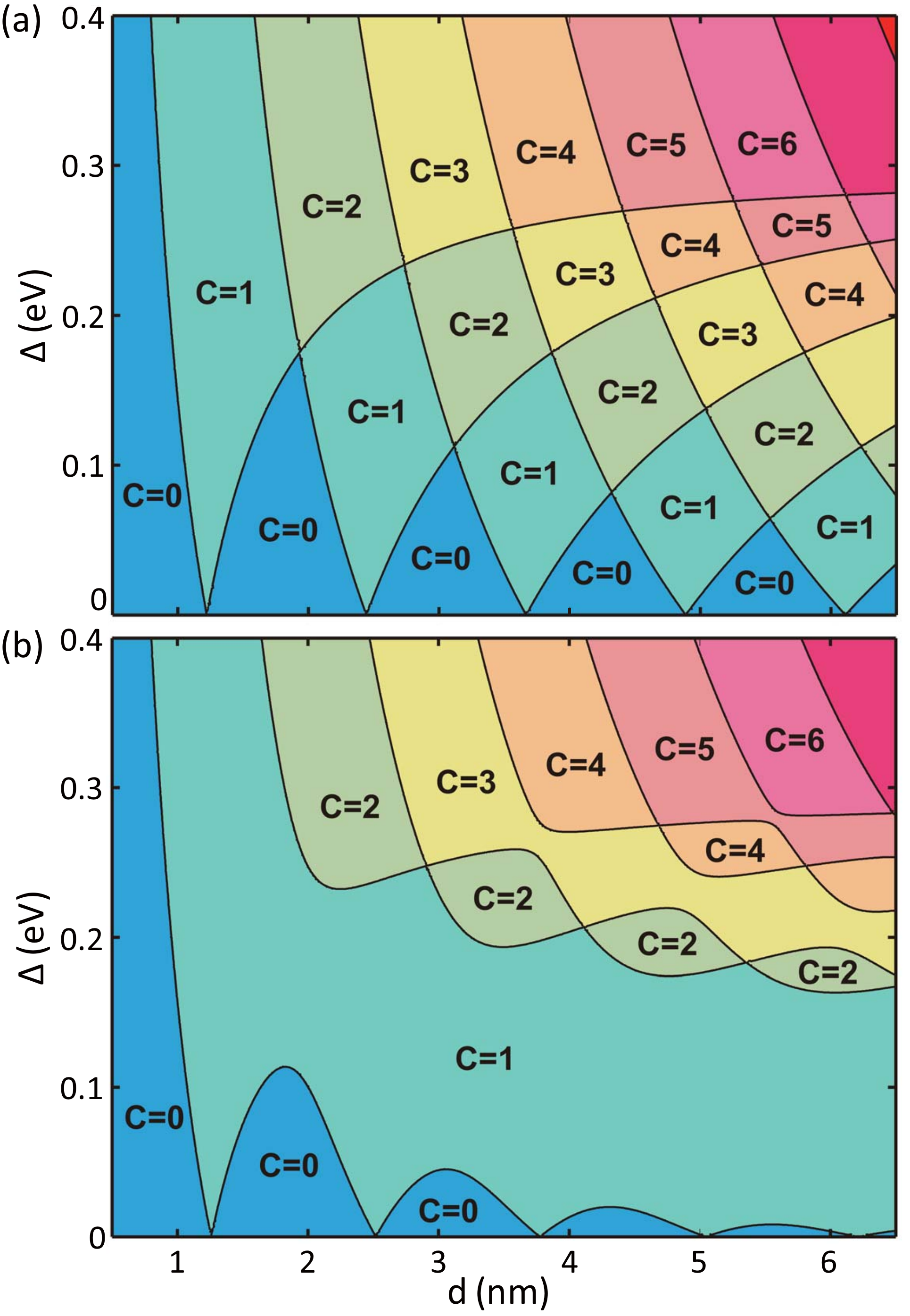}
\end{center}
\caption{The phase diagram of QAH effect in thin films of magnetic TIs with
two variables: the exchange field $\Delta$ and thickness of thin film $d$. All parameters are taken from Ref.~\cite{zhang2009} for Bi$_2$(Se$_{0.4}$Te$_{0.6}$)$_3$. (a) and (b) are phase diagrams without and with $A_1$ term, respectively. The different QAH phases are denoted by corresponding Chern numbers. The particle-hole asymmetric term is neglected for it does not change the topology of the phase diagram. The width of each Hall plateau in QAH effect depends on the band parameters and the thickness of the material, which is distinct from that in IQHE.} \label{fig2}
\end{figure}

With the criteria for the Chern number in Eq.~(\ref{chern}), we can identify a phase diagram in the parameter space ($\Delta, d$) as shown in Fig.~\ref{fig2}, where we adopt the parameters of Bi$_2$(Se$_{0.4}$Te$_{0.6}$)$_3$~\cite{zhang2009}, and neglect the particle-hole asymmetric term $\tilde{\varepsilon}_n$ for it does not change the condition for band inversion. In the absence of $A_1$ term, the condition for band inversion of $h_{\pm}(n)$ is $d>n\pi\sqrt{B_1/(\pm\Delta-B_0)}$, thus the phase boundaries are given by $d=n\pi\sqrt{B_1/(\pm\Delta-B_0)}$ [Fig.~\ref{fig2}(a)]. The different QAH phases are denoted by the corresponding Chern numbers. As shown in Fig.~\ref{fig2}(b), when $A_1$ term is turned on, it induces the coupling between $h_{\pm}(n)$ and $h_{\mp}(n+1)$, which makes the QAH phases with same Chern numbers simply connected in the phase diagram. Also it enlarges the $C=1$ phase and shrinks $C=2$ phase in the parameter space. The phase space of odd Chern number phases are simple connected, while those of even Chern number phases are separated into ``islands'', for the confinement potential has inversion symmetry along the $z$ direction.

For a given thickness, the Hall conductance experiences incremental plateaus $0$, $e^2/h$, $2e^2/h$,$\cdots$ as $\Delta$ increases. Remarkably, the inverse of the magnetization, proportional to $1/\Delta$ in QAH effect, is analogues to the magnetic field in IQHE. One the other hand, for a given exchange field, when the thickness $d$ is small enough, the band inversion in the bulk band structure will be removed  entirely by the finite size effect; with the increasing $d$, finite size effect is getting weaker and the band inversion among these sub-bands restores. If $\Delta>|B_0|$, the Dirac mass of class $h_-(n)$ can never be inverted, the Hall conductance plateaus transition always increase as $d$ increases. If $\Delta$ is small, the Dirac mass of the both classes can be inverted, and the system can only oscillate between $C=1$ QAH insulator and trivial band insulator as a function of thickness. Therefore, the QAH effect with higher plateaus requires a large enough exchange field.

The above discussion based on the analytic model give us a clear physical picture of the high Chern number QAH effect in magnetic TIs. In the following, we would like to consider about the possible realization in the realistic magnetic TI materials. The key point is to invert more spin polarized 2D sub-bands by large exchange field, while keeping the system full insulating. Thus the SOC of the materials has to be properly tuned to keep a full band gap. For Cr-doped (Bi,Sb)$_2$Te$_3$ magnetic TI used in experiment~\cite{chang2013b} to realize the $C=1$ QAH effect, the SOC of Te is so large that the $C=2$ QAH phase in this material becomes semi-metallic.

\begin{figure*}[t]
\begin{center}
\includegraphics[width=7.0in]{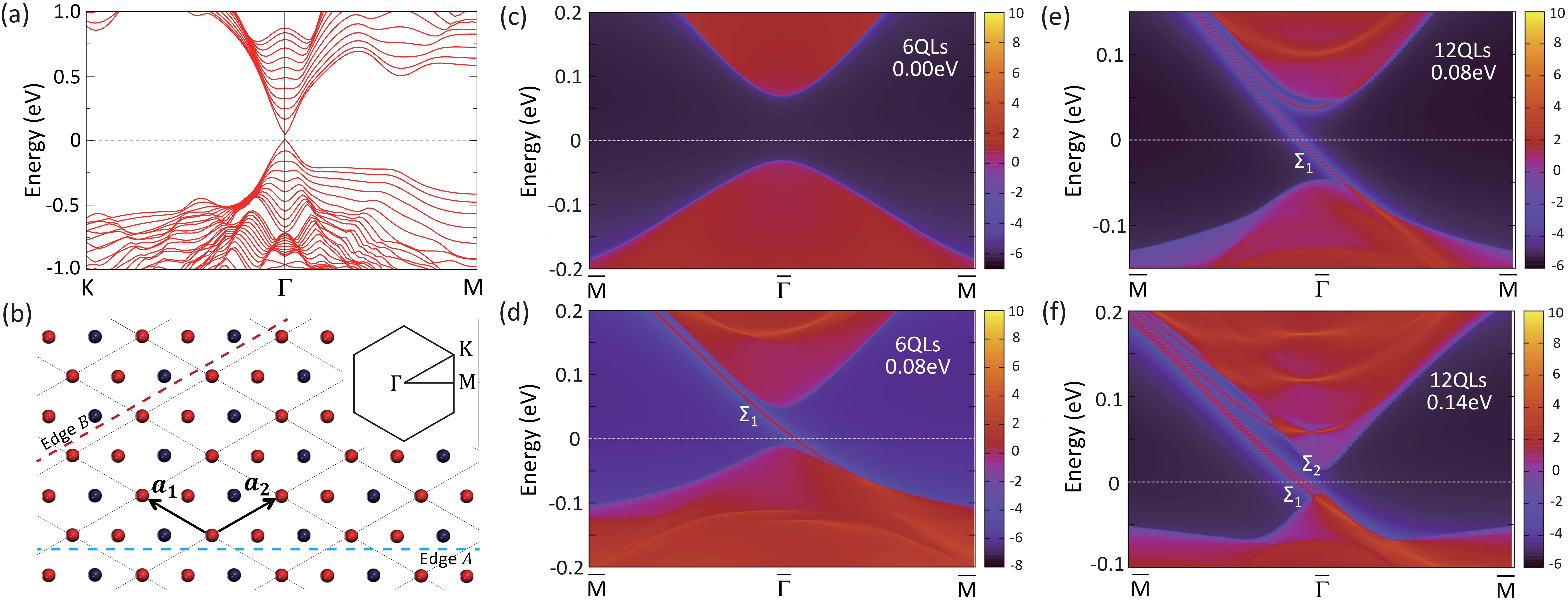}
\end{center}
\caption{Band structure, Brillouin zone and Edge states. (a) Band structure for 12-QLs Bi$_{1.78}$Cr$_{0.22}$(Se$_{0.4}$Te$_{0.6}$)$_3$ without exchange field. The dashed line indicates the Fermi level. (b) The top view of 2D thin film with two in-plane lattice vectors $\mathbf{a}_1$ and $\mathbf{a}_2$. The 1D edges are indicated by the dashed lines, \emph{Edge A} along $[1\bar{1}]$ direction (blue) and \emph{Edge B} along $[01]$ direction (red). The inset shows the 2D Brillouin zone, in which the high-symmetry \textbf{k} points $\Gamma$(0,0), K($\pi$,$\pi$) and M($\pi$,0) are labelled. (c)-(f) Energy and momentum dependence of the LDOS along \emph{Edge A} for the Bi$_{1.78}$Cr$_{0.22}$(Se$_{0.4}$Te$_{0.6}$)$_3$ film with thickness of 6-QLs and exchange field 0.0~eV (c), 0.08~eV (d) and thickness of 12-QLs and exchange field 0.08~eV (e), 0.14~eV (f). Here, the warmer colours represent higher LDOS. The red and blue regions indicate 2D bulk energy bands and energy gaps, respectively. The gapless chiral edge states can be clearly seen around the $\Gamma$ point as red lines dispersing in the 2D bulk gap. In (c), (d), (e), (f), the number of chiral edge state is $C=0, 1, 1, 2$.} \label{fig3}
\end{figure*}

We study the materials of Bi$_{2-y}$Cr$_y$(Se$_{x}$Te$_{1-x}$)$_3$ magnetic TI. With different Cr content and Se/Te ratio, the magnetic properties and SOC can be fine-tuned. We choose Cr-doped Bi$_2$(Se$_{0.4}$Te$_{0.6}$)$_3$ as an example, where  the Dirac cone of surface states is observed to locate in the bulk band gap~\cite{zhang2013}. Here, we first carried out the first-principles calculations on 3D Bi$_2$(Se$_{0.4}$Te$_{0.6}$)$_3$ without SOC, the virtual crystal approximation is employed to simulate the mixing between Se and Te in first-principles calculations. Then we get the effective SOC parameter of Bi$_{2-y}$Cr$_{y}$ by fitting the band structure of Bi$_{1.78}$Cr$_{0.22}$(Se$_{0.6}$Te$_{0.4}$)$_3$ in Ref.~\cite{zhang2013}. This system is at the critical point of the topological phase transition from inverted bands to normal bands, because the substitution of Bi by Cr reduces SOC strength. Finally, we construct the tight-binding model with SOC and the exchange interaction based on maximally localized Wannier functions~\cite{marzari1997,souza2001}. When the 2D system stays in the QAH phase, there are topologically protected chiral edge states at the 1D edge. To show the topological feature more explicitly, we calculate the dispersion spectra of the chiral edge states directly. As examples, here we study the edge states of the 6-QLs and 12-QLs of Bi$_{2-y}$Cr$_y$(Se$_{0.4}$Te$_{0.6}$)$_3$ film along $[1\bar{1}]$ direction (\emph{Edge A} along $\Gamma$-$M$), as shown in Fig.~\ref{fig3}. For a semi-infinite system, combining the tight-binding model with the iterative method~\cite{sancho1984}, we can calculate the Green's function for the edge states directly. The local density of states (LDOS) is directly related to the imaginary part of Green's function, from which we can obtain the dispersion of the edge states. As shown in Fig.~\ref{fig3}(f) for 12-QLs Bi$_{1.78}$Cr$_{0.22}$(Se$_{0.4}$Te$_{0.6}$)$_3$ with $\Delta=0.14$~eV, there indeed exist two gapless chiral edge states $\Sigma_1$ and $\Sigma_2$ in the 2D bulk gap indicating the $C=2$ QAH effect.

Recent experiments have shown that the thickness of thin films TIs can be well controlled through layer-by-layer growth via molecular beam epitaxy~\cite{zhang2010}, and the exchange field $\Delta$ can be tuned by changing the doping concentration $y$ of the magnetic elements~\cite{kulbachinskii2001,zhou2006,chang2013a}. In the mean field approximation, $\Delta$ can be estimated as $\Delta=yJ_{\mathrm{eff}}\langle S\rangle/2$, where $\langle S\rangle$ is the mean field expectation value of the local spin, and $J_{\mathrm{eff}}$ is the effective exchange parameter between local moments and the band electron. For Cr-doped Bi$_2$(Se,Te)$_3$, $\langle S\rangle=3/2$, $J_{\mathrm{eff}}$ is around $2.7$~eV~\cite{yu2010}, and the FM Curie temperature is about tens of K. With the concentration of the magnetic dopants to be $10\%$, the exchange field can be as large as $0.2$~eV, making the realization of QAH effect with higher plateaus feasible.

Experimentally, for the QAH effect with a higher Chern number $C$, the gate-tuned Hall resistance $\rho_{xy}$ should be accurately quantized into  $h/Ce^2$ plateau at zero magnetic field accompanied by a vanishing longitudinal resistance $\rho_{xx}$ and conductance as shown in Fig.~\ref{fig1}(c). In real materials, there always exists residual dissipative conduction channels contributed by a small amount of bulk carriers; however, if the carrier density is low enough, they will become localized states by the disorder potentials and will not affect the exact quantization of the Hall plateau.

The QAH effect with higher plateaus may provide a setting for both fundamental and applied investigation. A wealth of materials with tunable magnetic and topological properties~\cite{xu2011,brahlek2012} could lead to the discovery of more high Chern number QAH insulators. The multiple dissipationless edge channels in higher plateaus QAH effect would offer better ways to optimize electrical transport properties, leading to novel designs for low-power-consumption electronics.

We are grateful to X. L. Qi for insightful discussions. This work is supported
by the Defense Advanced Research Projects Agency Microsystems
Technology Office, MesoDynamic Architecture Program (MESO) through the contract number
N66001-11-1-4105, the DARPA Program on
"Topological Insulators -- Solid State Chemistry, New Materials and Properties".
under the award number N66001-12-1-4034 and by the US Department of
Energy, Office of Basic Energy Sciences, Division of Materials Sciences and Engineering, under contract
DE-AC02-76SF00515.


\begin{thebibliography}{25}%
\makeatletter
\providecommand \@ifxundefined [1]{%
 \@ifx{#1\undefined}
}%
\providecommand \@ifnum [1]{%
 \ifnum #1\expandafter \@firstoftwo
 \else \expandafter \@secondoftwo
 \fi
}%
\providecommand \@ifx [1]{%
 \ifx #1\expandafter \@firstoftwo
 \else \expandafter \@secondoftwo
 \fi
}%
\providecommand \natexlab [1]{#1}%
\providecommand \enquote  [1]{``#1''}%
\providecommand \bibnamefont  [1]{#1}%
\providecommand \bibfnamefont [1]{#1}%
\providecommand \citenamefont [1]{#1}%
\providecommand \href@noop [0]{\@secondoftwo}%
\providecommand \href [0]{\begingroup \@sanitize@url \@href}%
\providecommand \@href[1]{\@@startlink{#1}\@@href}%
\providecommand \@@href[1]{\endgroup#1\@@endlink}%
\providecommand \@sanitize@url [0]{\catcode `\\12\catcode `\$12\catcode
  `\&12\catcode `\#12\catcode `\^12\catcode `\_12\catcode `\%12\relax}%
\providecommand \@@startlink[1]{}%
\providecommand \@@endlink[0]{}%
\providecommand \url  [0]{\begingroup\@sanitize@url \@url }%
\providecommand \@url [1]{\endgroup\@href {#1}{\urlprefix }}%
\providecommand \urlprefix  [0]{URL }%
\providecommand \Eprint [0]{\href }%
\providecommand \doibase [0]{http://dx.doi.org/}%
\providecommand \selectlanguage [0]{\@gobble}%
\providecommand \bibinfo  [0]{\@secondoftwo}%
\providecommand \bibfield  [0]{\@secondoftwo}%
\providecommand \translation [1]{[#1]}%
\providecommand \BibitemOpen [0]{}%
\providecommand \bibitemStop [0]{}%
\providecommand \bibitemNoStop [0]{.\EOS\space}%
\providecommand \EOS [0]{\spacefactor3000\relax}%
\providecommand \BibitemShut  [1]{\csname bibitem#1\endcsname}%
\let\auto@bib@innerbib\@empty
\bibitem [{\citenamefont {Thouless}\ \emph {et~al.}(1982)\citenamefont
  {Thouless}, \citenamefont {Kohmoto}, \citenamefont {Nightingale},\ and\
  \citenamefont {den Nijs}}]{thouless1982}%
  \BibitemOpen
  \bibfield  {author} {\bibinfo {author} {\bibfnamefont {D.~J.}\ \bibnamefont
  {Thouless}}, \bibinfo {author} {\bibfnamefont {M.}~\bibnamefont {Kohmoto}},
  \bibinfo {author} {\bibfnamefont {M.~P.}\ \bibnamefont {Nightingale}}, \ and\
  \bibinfo {author} {\bibfnamefont {M.}~\bibnamefont {den Nijs}},\ }\href
  {\doibase 10.1103/PhysRevLett.49.405} {\bibfield  {journal} {\bibinfo
  {journal} {Phys. Rev. Lett.}\ }\textbf {\bibinfo {volume} {49}},\ \bibinfo
  {pages} {405} (\bibinfo {year} {1982})}\BibitemShut {NoStop}%
\bibitem [{\citenamefont {Klitzing}\ \emph {et~al.}(1980)\citenamefont
  {Klitzing}, \citenamefont {Dorda},\ and\ \citenamefont
  {Pepper}}]{klitzing1980}%
  \BibitemOpen
  \bibfield  {author} {\bibinfo {author} {\bibfnamefont {K.~v.}\ \bibnamefont
  {Klitzing}}, \bibinfo {author} {\bibfnamefont {G.}~\bibnamefont {Dorda}}, \
  and\ \bibinfo {author} {\bibfnamefont {M.}~\bibnamefont {Pepper}},\ }\href
  {\doibase 10.1103/PhysRevLett.45.494} {\bibfield  {journal} {\bibinfo
  {journal} {Phys. Rev. Lett.}\ }\textbf {\bibinfo {volume} {45}},\ \bibinfo
  {pages} {494} (\bibinfo {year} {1980})}\BibitemShut {NoStop}%
\bibitem [{\citenamefont {Halperin}(1982)}]{halperin1982}%
  \BibitemOpen
  \bibfield  {author} {\bibinfo {author} {\bibfnamefont {B.~I.}\ \bibnamefont
  {Halperin}},\ }\href {\doibase 10.1103/PhysRevB.25.2185} {\bibfield
  {journal} {\bibinfo  {journal} {Phys. Rev. B}\ }\textbf {\bibinfo {volume}
  {25}},\ \bibinfo {pages} {2185} (\bibinfo {year} {1982})}\BibitemShut
  {NoStop}%
\bibitem [{\citenamefont {Haldane}(1988)}]{haldane1988}%
  \BibitemOpen
  \bibfield  {author} {\bibinfo {author} {\bibfnamefont {F.~D.~M.}\
  \bibnamefont {Haldane}},\ }\href {\doibase 10.1103/PhysRevLett.61.2015}
  {\bibfield  {journal} {\bibinfo  {journal} {Phys. Rev. Lett.}\ }\textbf
  {\bibinfo {volume} {61}},\ \bibinfo {pages} {2015} (\bibinfo {year}
  {1988})}\BibitemShut {NoStop}%
\bibitem [{\citenamefont {Qi}\ \emph {et~al.}(2006)\citenamefont {Qi},
  \citenamefont {Wu},\ and\ \citenamefont {Zhang}}]{qi2006}%
  \BibitemOpen
  \bibfield  {author} {\bibinfo {author} {\bibfnamefont {X.-L.}\ \bibnamefont
  {Qi}}, \bibinfo {author} {\bibfnamefont {Y.-S.}\ \bibnamefont {Wu}}, \ and\
  \bibinfo {author} {\bibfnamefont {S.-C.}\ \bibnamefont {Zhang}},\ }\href
  {\doibase 10.1103/PhysRevB.74.085308} {\bibfield  {journal} {\bibinfo
  {journal} {Phys. Rev. B}\ }\textbf {\bibinfo {volume} {74}},\ \bibinfo
  {pages} {085308} (\bibinfo {year} {2006})}\BibitemShut {NoStop}%
\bibitem [{\citenamefont {Qi}\ \emph {et~al.}(2008)\citenamefont {Qi},
  \citenamefont {Hughes},\ and\ \citenamefont {Zhang}}]{qi2008}%
  \BibitemOpen
  \bibfield  {author} {\bibinfo {author} {\bibfnamefont {X.-L.}\ \bibnamefont
  {Qi}}, \bibinfo {author} {\bibfnamefont {T.~L.}\ \bibnamefont {Hughes}}, \
  and\ \bibinfo {author} {\bibfnamefont {S.-C.}\ \bibnamefont {Zhang}},\ }\href
  {\doibase 10.1103/PhysRevB.78.195424} {\bibfield  {journal} {\bibinfo
  {journal} {Phys. Rev. B}\ }\textbf {\bibinfo {volume} {78}},\ \bibinfo
  {pages} {195424} (\bibinfo {year} {2008})}\BibitemShut {NoStop}%
\bibitem [{\citenamefont {Liu}\ \emph {et~al.}(2008)\citenamefont {Liu},
  \citenamefont {Qi}, \citenamefont {Dai}, \citenamefont {Fang},\ and\
  \citenamefont {Zhang}}]{liu2008}%
  \BibitemOpen
  \bibfield  {author} {\bibinfo {author} {\bibfnamefont {C.-X.}\ \bibnamefont
  {Liu}}, \bibinfo {author} {\bibfnamefont {X.-L.}\ \bibnamefont {Qi}},
  \bibinfo {author} {\bibfnamefont {X.}~\bibnamefont {Dai}}, \bibinfo {author}
  {\bibfnamefont {Z.}~\bibnamefont {Fang}}, \ and\ \bibinfo {author}
  {\bibfnamefont {S.-C.}\ \bibnamefont {Zhang}},\ }\href {\doibase
  10.1103/PhysRevLett.101.146802} {\bibfield  {journal} {\bibinfo  {journal}
  {Phys. Rev. Lett.}\ }\textbf {\bibinfo {volume} {101}},\ \bibinfo {pages}
  {146802} (\bibinfo {year} {2008})}\BibitemShut {NoStop}%
\bibitem [{\citenamefont {Li}\ \emph {et~al.}(2010)\citenamefont {Li},
  \citenamefont {Wang}, \citenamefont {Qi},\ and\ \citenamefont
  {Zhang}}]{li2010}%
  \BibitemOpen
  \bibfield  {author} {\bibinfo {author} {\bibfnamefont {R.}~\bibnamefont
  {Li}}, \bibinfo {author} {\bibfnamefont {J.}~\bibnamefont {Wang}}, \bibinfo
  {author} {\bibfnamefont {X.~L.}\ \bibnamefont {Qi}}, \ and\ \bibinfo {author}
  {\bibfnamefont {S.~C.}\ \bibnamefont {Zhang}},\ }\href@noop {} {\bibfield
  {journal} {\bibinfo  {journal} {Nature Phys.}\ }\textbf {\bibinfo {volume}
  {6}},\ \bibinfo {pages} {284} (\bibinfo {year} {2010})}\BibitemShut {NoStop}%
\bibitem [{\citenamefont {Yu}\ \emph {et~al.}(2010)\citenamefont {Yu},
  \citenamefont {Zhang}, \citenamefont {Zhang}, \citenamefont {Zhang},
  \citenamefont {Dai},\ and\ \citenamefont {Fang}}]{yu2010}%
  \BibitemOpen
  \bibfield  {author} {\bibinfo {author} {\bibfnamefont {R.}~\bibnamefont
  {Yu}}, \bibinfo {author} {\bibfnamefont {W.}~\bibnamefont {Zhang}}, \bibinfo
  {author} {\bibfnamefont {H.-J.}\ \bibnamefont {Zhang}}, \bibinfo {author}
  {\bibfnamefont {S.-C.}\ \bibnamefont {Zhang}}, \bibinfo {author}
  {\bibfnamefont {X.}~\bibnamefont {Dai}}, \ and\ \bibinfo {author}
  {\bibfnamefont {Z.}~\bibnamefont {Fang}},\ }\href {\doibase
  10.1126/science.1187485} {\bibfield  {journal} {\bibinfo  {journal}
  {Science}\ }\textbf {\bibinfo {volume} {329}},\ \bibinfo {pages} {61}
  (\bibinfo {year} {2010})}\BibitemShut {NoStop}%
\bibitem [{\citenamefont {Chang}\ \emph
  {et~al.}(2013{\natexlab{a}})\citenamefont {Chang}, \citenamefont {Zhang},
  \citenamefont {Feng}, \citenamefont {Shen}, \citenamefont {Zhang},
  \citenamefont {Guo}, \citenamefont {Li}, \citenamefont {Ou}, \citenamefont
  {Wei}, \citenamefont {Wang}, \citenamefont {Ji}, \citenamefont {Feng},
  \citenamefont {Ji}, \citenamefont {Chen}, \citenamefont {Jia}, \citenamefont
  {Dai}, \citenamefont {Fang}, \citenamefont {Zhang}, \citenamefont {He},
  \citenamefont {Wang}, \citenamefont {Lu}, \citenamefont {Ma},\ and\
  \citenamefont {Xue}}]{chang2013b}%
  \BibitemOpen
  \bibfield  {author} {\bibinfo {author} {\bibfnamefont {C.-Z.}\ \bibnamefont
  {Chang}}, \bibinfo {author} {\bibfnamefont {J.}~\bibnamefont {Zhang}},
  \bibinfo {author} {\bibfnamefont {X.}~\bibnamefont {Feng}}, \bibinfo {author}
  {\bibfnamefont {J.}~\bibnamefont {Shen}}, \bibinfo {author} {\bibfnamefont
  {Z.}~\bibnamefont {Zhang}}, \bibinfo {author} {\bibfnamefont
  {M.}~\bibnamefont {Guo}}, \bibinfo {author} {\bibfnamefont {K.}~\bibnamefont
  {Li}}, \bibinfo {author} {\bibfnamefont {Y.}~\bibnamefont {Ou}}, \bibinfo
  {author} {\bibfnamefont {P.}~\bibnamefont {Wei}}, \bibinfo {author}
  {\bibfnamefont {L.-L.}\ \bibnamefont {Wang}}, \bibinfo {author}
  {\bibfnamefont {Z.-Q.}\ \bibnamefont {Ji}}, \bibinfo {author} {\bibfnamefont
  {Y.}~\bibnamefont {Feng}}, \bibinfo {author} {\bibfnamefont {S.}~\bibnamefont
  {Ji}}, \bibinfo {author} {\bibfnamefont {X.}~\bibnamefont {Chen}}, \bibinfo
  {author} {\bibfnamefont {J.}~\bibnamefont {Jia}}, \bibinfo {author}
  {\bibfnamefont {X.}~\bibnamefont {Dai}}, \bibinfo {author} {\bibfnamefont
  {Z.}~\bibnamefont {Fang}}, \bibinfo {author} {\bibfnamefont {S.-C.}\
  \bibnamefont {Zhang}}, \bibinfo {author} {\bibfnamefont {K.}~\bibnamefont
  {He}}, \bibinfo {author} {\bibfnamefont {Y.}~\bibnamefont {Wang}}, \bibinfo
  {author} {\bibfnamefont {L.}~\bibnamefont {Lu}}, \bibinfo {author}
  {\bibfnamefont {X.-C.}\ \bibnamefont {Ma}}, \ and\ \bibinfo {author}
  {\bibfnamefont {Q.-K.}\ \bibnamefont {Xue}},\ }\href {\doibase
  10.1126/science.1234414} {\bibfield  {journal} {\bibinfo  {journal}
  {Science}\ }\textbf {\bibinfo {volume} {340}},\ \bibinfo {pages} {167}
  (\bibinfo {year} {2013}{\natexlab{a}})}\BibitemShut {NoStop}%
\bibitem [{\citenamefont {Zhang}\ and\ \citenamefont
  {Zhang}(2012)}]{zhang2012}%
  \BibitemOpen
  \bibfield  {author} {\bibinfo {author} {\bibfnamefont {X.}~\bibnamefont
  {Zhang}}\ and\ \bibinfo {author} {\bibfnamefont {S.-C.}\ \bibnamefont
  {Zhang}},\ }\href@noop {} {\bibfield  {journal} {\bibinfo  {journal} {MICRO-
  AND NANOTECHNOLOGY SENSORS, SYSTEMS, AND APPLICATIONS IV, Proceedings of
  SPIE}\ }\textbf {\bibinfo {volume} {8373}},\ \bibinfo {pages} {837309}
  (\bibinfo {year} {2012})}\BibitemShut {NoStop}%
\bibitem [{\citenamefont {Barkeshli}\ and\ \citenamefont
  {Qi}(2012)}]{maissam2012}%
  \BibitemOpen
  \bibfield  {author} {\bibinfo {author} {\bibfnamefont {M.}~\bibnamefont
  {Barkeshli}}\ and\ \bibinfo {author} {\bibfnamefont {X.-L.}\ \bibnamefont
  {Qi}},\ }\href {\doibase 10.1103/PhysRevX.2.031013} {\bibfield  {journal}
  {\bibinfo  {journal} {Phys. Rev. X}\ }\textbf {\bibinfo {volume} {2}},\
  \bibinfo {pages} {031013} (\bibinfo {year} {2012})}\BibitemShut {NoStop}%
\bibitem [{\citenamefont {Bernevig}\ \emph {et~al.}(2006)\citenamefont
  {Bernevig}, \citenamefont {Hughes},\ and\ \citenamefont
  {Zhang}}]{bernevig2006c}%
  \BibitemOpen
  \bibfield  {author} {\bibinfo {author} {\bibfnamefont {B.~A.}\ \bibnamefont
  {Bernevig}}, \bibinfo {author} {\bibfnamefont {T.~L.}\ \bibnamefont
  {Hughes}}, \ and\ \bibinfo {author} {\bibfnamefont {S.~C.}\ \bibnamefont
  {Zhang}},\ }\href@noop {} {\bibfield  {journal} {\bibinfo  {journal}
  {Science}\ }\textbf {\bibinfo {volume} {314}},\ \bibinfo {pages} {1757}
  (\bibinfo {year} {2006})}\BibitemShut {NoStop}%
\bibitem [{\citenamefont {Zhang}\ \emph {et~al.}(2009)\citenamefont {Zhang},
  \citenamefont {Liu}, \citenamefont {Qi}, \citenamefont {Dai}, \citenamefont
  {Fang},\ and\ \citenamefont {Zhang}}]{zhang2009}%
  \BibitemOpen
  \bibfield  {author} {\bibinfo {author} {\bibfnamefont {H.}~\bibnamefont
  {Zhang}}, \bibinfo {author} {\bibfnamefont {C.-X.}\ \bibnamefont {Liu}},
  \bibinfo {author} {\bibfnamefont {X.-L.}\ \bibnamefont {Qi}}, \bibinfo
  {author} {\bibfnamefont {X.}~\bibnamefont {Dai}}, \bibinfo {author}
  {\bibfnamefont {Z.}~\bibnamefont {Fang}}, \ and\ \bibinfo {author}
  {\bibfnamefont {S.-C.}\ \bibnamefont {Zhang}},\ }\href@noop {} {\bibfield
  {journal} {\bibinfo  {journal} {Nature Phys.}\ }\textbf {\bibinfo {volume}
  {5}},\ \bibinfo {pages} {438} (\bibinfo {year} {2009})}\BibitemShut {NoStop}%
\bibitem [{\citenamefont {Kul'bachinskii}\ \emph {et~al.}(2001)\citenamefont
  {Kul'bachinskii}, \citenamefont {Kaminskii}, \citenamefont {Kindo},
  \citenamefont {Narumi}, \citenamefont {Suga}, \citenamefont {Lostak},\ and\
  \citenamefont {Svanda}}]{kulbachinskii2001}%
  \BibitemOpen
  \bibfield  {author} {\bibinfo {author} {\bibfnamefont {V.}~\bibnamefont
  {Kul'bachinskii}}, \bibinfo {author} {\bibfnamefont {A.}~\bibnamefont
  {Kaminskii}}, \bibinfo {author} {\bibfnamefont {K.}~\bibnamefont {Kindo}},
  \bibinfo {author} {\bibfnamefont {Y.}~\bibnamefont {Narumi}}, \bibinfo
  {author} {\bibfnamefont {K.}~\bibnamefont {Suga}}, \bibinfo {author}
  {\bibfnamefont {P.}~\bibnamefont {Lostak}}, \ and\ \bibinfo {author}
  {\bibfnamefont {P.}~\bibnamefont {Svanda}},\ }\href {\doibase
  10.1134/1.1378118} {\bibfield  {journal} {\bibinfo  {journal} {JETP Lett.}\
  }\textbf {\bibinfo {volume} {73}},\ \bibinfo {pages} {352} (\bibinfo {year}
  {2001})}\BibitemShut {NoStop}%
\bibitem [{\citenamefont {Zhou}\ \emph {et~al.}(2006)\citenamefont {Zhou},
  \citenamefont {Chien},\ and\ \citenamefont {Uher}}]{zhou2006}%
  \BibitemOpen
  \bibfield  {author} {\bibinfo {author} {\bibfnamefont {Z.}~\bibnamefont
  {Zhou}}, \bibinfo {author} {\bibfnamefont {Y.-J.}\ \bibnamefont {Chien}}, \
  and\ \bibinfo {author} {\bibfnamefont {C.}~\bibnamefont {Uher}},\ }\href
  {\doibase 10.1103/PhysRevB.74.224418} {\bibfield  {journal} {\bibinfo
  {journal} {Phys. Rev. B}\ }\textbf {\bibinfo {volume} {74}},\ \bibinfo
  {pages} {224418} (\bibinfo {year} {2006})}\BibitemShut {NoStop}%
\bibitem [{\citenamefont {Chang}\ \emph
  {et~al.}(2013{\natexlab{b}})\citenamefont {Chang}, \citenamefont {Zhang},
  \citenamefont {Liu}, \citenamefont {Zhang}, \citenamefont {Feng},
  \citenamefont {Li}, \citenamefont {Wang}, \citenamefont {Chen}, \citenamefont
  {Dai}, \citenamefont {Fang}, \citenamefont {Qi}, \citenamefont {Zhang},
  \citenamefont {Wang}, \citenamefont {He}, \citenamefont {Ma},\ and\
  \citenamefont {Xue}}]{chang2013a}%
  \BibitemOpen
  \bibfield  {author} {\bibinfo {author} {\bibfnamefont {C.-Z.}\ \bibnamefont
  {Chang}}, \bibinfo {author} {\bibfnamefont {J.}~\bibnamefont {Zhang}},
  \bibinfo {author} {\bibfnamefont {M.}~\bibnamefont {Liu}}, \bibinfo {author}
  {\bibfnamefont {Z.}~\bibnamefont {Zhang}}, \bibinfo {author} {\bibfnamefont
  {X.}~\bibnamefont {Feng}}, \bibinfo {author} {\bibfnamefont {K.}~\bibnamefont
  {Li}}, \bibinfo {author} {\bibfnamefont {L.-L.}\ \bibnamefont {Wang}},
  \bibinfo {author} {\bibfnamefont {X.}~\bibnamefont {Chen}}, \bibinfo {author}
  {\bibfnamefont {X.}~\bibnamefont {Dai}}, \bibinfo {author} {\bibfnamefont
  {Z.}~\bibnamefont {Fang}}, \bibinfo {author} {\bibfnamefont {X.-L.}\
  \bibnamefont {Qi}}, \bibinfo {author} {\bibfnamefont {S.-C.}\ \bibnamefont
  {Zhang}}, \bibinfo {author} {\bibfnamefont {Y.}~\bibnamefont {Wang}},
  \bibinfo {author} {\bibfnamefont {K.}~\bibnamefont {He}}, \bibinfo {author}
  {\bibfnamefont {X.-C.}\ \bibnamefont {Ma}}, \ and\ \bibinfo {author}
  {\bibfnamefont {Q.-K.}\ \bibnamefont {Xue}},\ }\href {\doibase
  10.1002/adma.201203493} {\bibfield  {journal} {\bibinfo  {journal} {Adv.
  Mater.}\ }\textbf {\bibinfo {volume} {25}},\ \bibinfo {pages} {1065}
  (\bibinfo {year} {2013}{\natexlab{b}})}\BibitemShut {NoStop}%
\bibitem [{\citenamefont {Liu}\ \emph {et~al.}(2010)\citenamefont {Liu},
  \citenamefont {Zhang}, \citenamefont {Yan}, \citenamefont {Qi}, \citenamefont
  {Frauenheim}, \citenamefont {Dai}, \citenamefont {Fang},\ and\ \citenamefont
  {Zhang}}]{liu2010a}%
  \BibitemOpen
  \bibfield  {author} {\bibinfo {author} {\bibfnamefont {C.-X.}\ \bibnamefont
  {Liu}}, \bibinfo {author} {\bibfnamefont {H.}~\bibnamefont {Zhang}}, \bibinfo
  {author} {\bibfnamefont {B.}~\bibnamefont {Yan}}, \bibinfo {author}
  {\bibfnamefont {X.-L.}\ \bibnamefont {Qi}}, \bibinfo {author} {\bibfnamefont
  {T.}~\bibnamefont {Frauenheim}}, \bibinfo {author} {\bibfnamefont
  {X.}~\bibnamefont {Dai}}, \bibinfo {author} {\bibfnamefont {Z.}~\bibnamefont
  {Fang}}, \ and\ \bibinfo {author} {\bibfnamefont {S.-C.}\ \bibnamefont
  {Zhang}},\ }\href {\doibase 10.1103/PhysRevB.81.041307} {\bibfield  {journal}
  {\bibinfo  {journal} {Phys. Rev. B}\ }\textbf {\bibinfo {volume} {81}},\
  \bibinfo {pages} {041307} (\bibinfo {year} {2010})}\BibitemShut {NoStop}%
\bibitem [{\citenamefont {Zhang}\ \emph {et~al.}(2013)\citenamefont {Zhang},
  \citenamefont {Chang}, \citenamefont {Tang}, \citenamefont {Zhang},
  \citenamefont {Feng}, \citenamefont {Li}, \citenamefont {Wang}, \citenamefont
  {Chen}, \citenamefont {Liu}, \citenamefont {Duan}, \citenamefont {He},
  \citenamefont {Xue}, \citenamefont {Ma},\ and\ \citenamefont
  {Wang}}]{zhang2013}%
  \BibitemOpen
  \bibfield  {author} {\bibinfo {author} {\bibfnamefont {J.}~\bibnamefont
  {Zhang}}, \bibinfo {author} {\bibfnamefont {C.-Z.}\ \bibnamefont {Chang}},
  \bibinfo {author} {\bibfnamefont {P.}~\bibnamefont {Tang}}, \bibinfo {author}
  {\bibfnamefont {Z.}~\bibnamefont {Zhang}}, \bibinfo {author} {\bibfnamefont
  {X.}~\bibnamefont {Feng}}, \bibinfo {author} {\bibfnamefont {K.}~\bibnamefont
  {Li}}, \bibinfo {author} {\bibfnamefont {L.-l.}\ \bibnamefont {Wang}},
  \bibinfo {author} {\bibfnamefont {X.}~\bibnamefont {Chen}}, \bibinfo {author}
  {\bibfnamefont {C.}~\bibnamefont {Liu}}, \bibinfo {author} {\bibfnamefont
  {W.}~\bibnamefont {Duan}}, \bibinfo {author} {\bibfnamefont {K.}~\bibnamefont
  {He}}, \bibinfo {author} {\bibfnamefont {Q.-K.}\ \bibnamefont {Xue}},
  \bibinfo {author} {\bibfnamefont {X.}~\bibnamefont {Ma}}, \ and\ \bibinfo
  {author} {\bibfnamefont {Y.}~\bibnamefont {Wang}},\ }\href {\doibase
  10.1126/science.1230905} {\bibfield  {journal} {\bibinfo  {journal}
  {Science}\ }\textbf {\bibinfo {volume} {339}},\ \bibinfo {pages} {1582}
  (\bibinfo {year} {2013})}\BibitemShut {NoStop}%
\bibitem [{\citenamefont {Marzari}\ and\ \citenamefont
  {Vanderbilt}(1997)}]{marzari1997}%
  \BibitemOpen
  \bibfield  {author} {\bibinfo {author} {\bibfnamefont {N.}~\bibnamefont
  {Marzari}}\ and\ \bibinfo {author} {\bibfnamefont {D.}~\bibnamefont
  {Vanderbilt}},\ }\href {\doibase 10.1103/PhysRevB.56.12847} {\bibfield
  {journal} {\bibinfo  {journal} {Phys. Rev. B}\ }\textbf {\bibinfo {volume}
  {56}},\ \bibinfo {pages} {12847} (\bibinfo {year} {1997})}\BibitemShut
  {NoStop}%
\bibitem [{\citenamefont {Souza}\ \emph {et~al.}(2001)\citenamefont {Souza},
  \citenamefont {Marzari},\ and\ \citenamefont {Vanderbilt}}]{souza2001}%
  \BibitemOpen
  \bibfield  {author} {\bibinfo {author} {\bibfnamefont {I.}~\bibnamefont
  {Souza}}, \bibinfo {author} {\bibfnamefont {N.}~\bibnamefont {Marzari}}, \
  and\ \bibinfo {author} {\bibfnamefont {D.}~\bibnamefont {Vanderbilt}},\
  }\href {\doibase 10.1103/PhysRevB.65.035109} {\bibfield  {journal} {\bibinfo
  {journal} {Phys. Rev. B}\ }\textbf {\bibinfo {volume} {65}},\ \bibinfo
  {pages} {035109} (\bibinfo {year} {2001})}\BibitemShut {NoStop}%
\bibitem [{\citenamefont {Sancho}\ \emph {et~al.}(1984)\citenamefont {Sancho},
  \citenamefont {Sancho},\ and\ \citenamefont {Rubio}}]{sancho1984}%
  \BibitemOpen
  \bibfield  {author} {\bibinfo {author} {\bibfnamefont {M.~P.~L.}\
  \bibnamefont {Sancho}}, \bibinfo {author} {\bibfnamefont {J.~M.~L.}\
  \bibnamefont {Sancho}}, \ and\ \bibinfo {author} {\bibfnamefont
  {J.}~\bibnamefont {Rubio}},\ }\href@noop {} {\bibfield  {journal} {\bibinfo
  {journal} {Journal of Physics F: Metal Physics}\ }\textbf {\bibinfo {volume}
  {14}},\ \bibinfo {pages} {1205} (\bibinfo {year} {1984})}\BibitemShut
  {NoStop}%
\bibitem [{\citenamefont {Zhang}\ \emph {et~al.}(2010)\citenamefont {Zhang},
  \citenamefont {He}, \citenamefont {Chang}, \citenamefont {Song},
  \citenamefont {Wang}, \citenamefont {Chen}, \citenamefont {Jia},
  \citenamefont {Fang}, \citenamefont {Dai}, \citenamefont {Shan},
  \citenamefont {Shen}, \citenamefont {Niu}, \citenamefont {Qi}, \citenamefont
  {Zhang}, \citenamefont {Ma},\ and\ \citenamefont {Xue}}]{zhang2010}%
  \BibitemOpen
  \bibfield  {author} {\bibinfo {author} {\bibfnamefont {Y.}~\bibnamefont
  {Zhang}}, \bibinfo {author} {\bibfnamefont {K.}~\bibnamefont {He}}, \bibinfo
  {author} {\bibfnamefont {C.-Z.}\ \bibnamefont {Chang}}, \bibinfo {author}
  {\bibfnamefont {C.-L.}\ \bibnamefont {Song}}, \bibinfo {author}
  {\bibfnamefont {L.-L.}\ \bibnamefont {Wang}}, \bibinfo {author}
  {\bibfnamefont {X.}~\bibnamefont {Chen}}, \bibinfo {author} {\bibfnamefont
  {J.-F.}\ \bibnamefont {Jia}}, \bibinfo {author} {\bibfnamefont
  {Z.}~\bibnamefont {Fang}}, \bibinfo {author} {\bibfnamefont {X.}~\bibnamefont
  {Dai}}, \bibinfo {author} {\bibfnamefont {W.-Y.}\ \bibnamefont {Shan}},
  \bibinfo {author} {\bibfnamefont {S.-Q.}\ \bibnamefont {Shen}}, \bibinfo
  {author} {\bibfnamefont {Q.}~\bibnamefont {Niu}}, \bibinfo {author}
  {\bibfnamefont {X.-L.}\ \bibnamefont {Qi}}, \bibinfo {author} {\bibfnamefont
  {S.-C.}\ \bibnamefont {Zhang}}, \bibinfo {author} {\bibfnamefont {X.-C.}\
  \bibnamefont {Ma}}, \ and\ \bibinfo {author} {\bibfnamefont {Q.-K.}\
  \bibnamefont {Xue}},\ }\href {\doibase 10.1038/nphys1689} {\bibfield
  {journal} {\bibinfo  {journal} {Nature Phys.}\ }\textbf {\bibinfo {volume}
  {6}},\ \bibinfo {pages} {584} (\bibinfo {year} {2010})}\BibitemShut {NoStop}%
\bibitem [{\citenamefont {Xu}\ \emph {et~al.}(2011)\citenamefont {Xu},
  \citenamefont {Xia}, \citenamefont {Wray}, \citenamefont {Jia}, \citenamefont
  {Meier}, \citenamefont {Dil}, \citenamefont {Osterwalder}, \citenamefont
  {Slomski}, \citenamefont {Bansil}, \citenamefont {Lin}, \citenamefont
  {Cava},\ and\ \citenamefont {Hasan}}]{xu2011}%
  \BibitemOpen
  \bibfield  {author} {\bibinfo {author} {\bibfnamefont {S.-Y.}\ \bibnamefont
  {Xu}}, \bibinfo {author} {\bibfnamefont {Y.}~\bibnamefont {Xia}}, \bibinfo
  {author} {\bibfnamefont {L.~A.}\ \bibnamefont {Wray}}, \bibinfo {author}
  {\bibfnamefont {S.}~\bibnamefont {Jia}}, \bibinfo {author} {\bibfnamefont
  {F.}~\bibnamefont {Meier}}, \bibinfo {author} {\bibfnamefont {J.~H.}\
  \bibnamefont {Dil}}, \bibinfo {author} {\bibfnamefont {J.}~\bibnamefont
  {Osterwalder}}, \bibinfo {author} {\bibfnamefont {B.}~\bibnamefont
  {Slomski}}, \bibinfo {author} {\bibfnamefont {A.}~\bibnamefont {Bansil}},
  \bibinfo {author} {\bibfnamefont {H.}~\bibnamefont {Lin}}, \bibinfo {author}
  {\bibfnamefont {R.~J.}\ \bibnamefont {Cava}}, \ and\ \bibinfo {author}
  {\bibfnamefont {M.~Z.}\ \bibnamefont {Hasan}},\ }\href {\doibase
  10.1126/science.1201607} {\bibfield  {journal} {\bibinfo  {journal}
  {Science}\ }\textbf {\bibinfo {volume} {332}},\ \bibinfo {pages} {560}
  (\bibinfo {year} {2011})}\BibitemShut {NoStop}%
\bibitem [{\citenamefont {Brahlek}\ \emph {et~al.}(2012)\citenamefont
  {Brahlek}, \citenamefont {Bansal}, \citenamefont {Koirala}, \citenamefont
  {Xu}, \citenamefont {Neupane}, \citenamefont {Liu}, \citenamefont {Hasan},\
  and\ \citenamefont {Oh}}]{brahlek2012}%
  \BibitemOpen
  \bibfield  {author} {\bibinfo {author} {\bibfnamefont {M.}~\bibnamefont
  {Brahlek}}, \bibinfo {author} {\bibfnamefont {N.}~\bibnamefont {Bansal}},
  \bibinfo {author} {\bibfnamefont {N.}~\bibnamefont {Koirala}}, \bibinfo
  {author} {\bibfnamefont {S.-Y.}\ \bibnamefont {Xu}}, \bibinfo {author}
  {\bibfnamefont {M.}~\bibnamefont {Neupane}}, \bibinfo {author} {\bibfnamefont
  {C.}~\bibnamefont {Liu}}, \bibinfo {author} {\bibfnamefont {M.~Z.}\
  \bibnamefont {Hasan}}, \ and\ \bibinfo {author} {\bibfnamefont
  {S.}~\bibnamefont {Oh}},\ }\href {\doibase 10.1103/PhysRevLett.109.186403}
  {\bibfield  {journal} {\bibinfo  {journal} {Phys. Rev. Lett.}\ }\textbf
  {\bibinfo {volume} {109}},\ \bibinfo {pages} {186403} (\bibinfo {year}
  {2012})}\BibitemShut {NoStop}%
\end{thebibliography}
\end{document}